# Advances in Precise Radial Velocimetry From Cross-Disciplinary Work in Heliophysics, Stellar Astronomy, and Instrumentation

Jason T. Wright
(814) 863-8470
Penn State University
astrowright@gmail.com

Steinn Sigurdsson
Penn State University

This version includes supplemental material added for the arXiv submission

# Advances in Precise Radial Velocimetry From Cross-Disciplinary Work in Heliophysics, Stellar Astronomy, and Instrumentation

Jason T. Wright and Steinn Sigurdsson

## A Successful Early Example of Cross-Disciplinary Collaboration

From August 28–September 18, 2016, The Aspen Center for Physics held a summer workshop entitled *Approaching the Stellar Astrophysical Limits of Exoplanet Detection: Getting to 10 cm/s.* The workshop brought together heliophysicists, stellar astrophysicists, and exoplanetary astronomers, including observers, theorists, and instrument builders to explore *fundamental* limits and challenges to precise stellar Doppler work. It was organized by Gibor Basri, Fabienne Bastien, Xavier Dumusque, and Jason T. Wright.

The workshop participants agreed on several important outcomes from the workshop. In particular:

- Collaboration *among* precise RV groups is vital. Having multiple instruments observe a common set of bright radial velocity "quiet" stars at high cadence will generate the data set the community needs to crack the problem, and also help separate idiosyncratic instrumental effects from stellar variations.

- Collaboration *beyond* the precise RV community is vital. Collaboration with the heliophysics community will prevent the stellar community from spending time re-discovering well understood phenomena in stellar photospheres, and interaction with other hardware communities will accelerate the development of instrumental advances.

- The heliophysics community has much to offer the stellar community, but significant barriers exist preventing a fruitful exchange of knowledge including:
    - Few opportunities for interaction
    - Significant jargon differences
    - Insufficient appreciation on both sides for opportunities for funding
    - Insufficient appreciation on both sides for opportunities for scientific advances

A few potentially promising avenues for collaboration that could bear significant fruit come from the solar 3D magnetohydrodynamic modeling community, for example:

## One Example: Synthetic Stellar Spectra as a Map of RV Jitter Diagnostics

Stellar astronomers work entirely with *disk-integrated spectra*, while heliophysicists are most often concerned with individual regions of the Sun. Heliophysicists thus often have detailed understandings of the small-scale processes, but spend much less effort on understanding the globally averaged phenomena that stellar astronomers study. The application of 3D solar photospheric modeling codes to generate as-observed stellar photospheric spectra is difficult and computationally intensive, but also practical and could potentially generate significant scientific return.

Analysis of such synthetic spectra might then reveal which lines or spectral features are *diagnostic* of non-center-of-mass Doppler shifts (i.e. stellar "jitter" or photospheric "noise"). At the moment, stellar astronomers must average the Doppler information of the many lines in a stellar spectrum, and generate statistical weights for these lines that produce the "best" precise radial velocity measurements. This purely empirical approach is challenging because the sources of radial velocity "jitter" at the 1 m/s level and below are still not well understood, and likely have multiple causes. While it is clear that spots, magnetic suppression of convection, p-mode oscillations, and granulation all contribute to this RV "noise," their relative contributions are still not clear.

The generation of diagnostics for some or all forms of RV jitter—for instance, the understanding that a particular profile change of a particularly sensitive set of lines is indicative of a global shift due to the suppression of convective blueshift in plage regions—would allow for the "correction" of measured RVs using additional data from the very spectra used to measure the bulk RV variation, and the development of instruments and observing techniques for ensuring that these diagnostics are well measured.

Such a project might generate significant scientific returns, but would be challenging to undertake. Even beyond the difficulty of modeling an *entire* stellar disk and modifying code designed for the Sun to appropriately model *other* stars, modelers often must make significant compromises for the sake of computability. It is therefore essential that the RV community be explicit about what they need and to what fidelity they need it, since the needs of the two communities tend to be very different in non-obvious ways. For instance, the importance of non-LTE effects and magnetic fields on precise stellar Doppler work can be crucial in many cases, and irrelevant in others. Since including these effects can significantly burden calculations, stellar astronomers would have to work closely with modelers in determining which effects to include and which can safely be avoided.

**Further Synthesis**

Some additional opportunities for collaboration among the communities:

Observations of the Sun-as-a-star have a long history, and recent advances in hardware have allowed the Sun to be regularly observed as a precise radial velocity target (for instance at HARPS-N: Haywood et al. 2016 Cool Stars 19, 47; Phillips et al. 2016, SPIE 99126Z). This allows for a direct comparison of heliophysics data sets, including Dopplergrams and magnetograms, with stellar spectra in the same instrument used to measure stellar radial velocities.

Many assumptions in the RV community, for instance the utility of broad wavelength coverage in diagnosing non-center-of-mass Doppler shifts, the use of line bisectors as a primary diagnostic, and the importance of spot modeling, were challenged by the heliophysicists at the workshop as not obviously valid or necessarily robust. This "outsiders'" perspective of the field should feed back into the design requirements of the next generation of instruments, including parameters such as wavelength coverage, resolution, throughput, dispersion, and polarimetric modes.

## A Bright Future

The Aspen Workshop demonstrated that the problem of the detection of true Earth analogs can best be addressed through a collaborative effort. It is unlikely that any single precise RV group will "crack the nut" of stellar jitter with one especially good instrument. It is likely that the collaboration that will solve the problem will be international, open, and cross-disciplinary, and work because of close collaboration among statisticians, heliophysicists, instrument builders, and stellar astronomers.

It is clear from the workshop that a key question that will remain after current and planned exoplanet missions are completed is precisely how to "de-jitter" precise radial velocities, and detect planets with RV amplitudes significantly below the stellar jitter "floor" near 1 m/s. Key challenges include: the theoretical understanding of which photospheric processes dominate RV jitter in a particular star, the observational identification of spectral diagnostics of non-center-of-mass Doppler shifts, and the computational challenge of creating a synthetic disk-integrated spectrum that connects these theoretical predictions to actual observed spectral features.

To make progress on these challenges, the stellar, exoplanetary, and instrument scientists in the precise RV community must be better engaged with the heliophysics community. This will require funding for collaboration across these disciplines, dedicated meetings where the communities can learn about each other and interact in person, and computational and observational resources to generate the appropriate synthetic and real data sets to drive and prove theoretical understanding of RV jitter.

Such cross-disciplinary work will accelerate progress in future areas of exoplanetary discovery, including, ultimately the RV detection and characterization of true Earth analogs, allowing their masses (and, so compositions) to be measured and understood.

# Supplemental material for arXiv version

This white paper was submitted to the National Academy of Sciences, Engineering, and Medicine Committee on Exoplanet Science Strategy.

The committee's final report, Exoplanet Science Strategy (National Academies 2018), cites this white paper twice:

> *The standard approach of tracking stellar activity using familiar activity indicators (e.g., changes in the equivalent widths of the Ca H and K lines) has generally proven to have limited value. New methods should be developed that attempt to empirically model and correct for stellar variability in time. Theorists studying magneto- hydrodynamics, stellar activity, stellar astrophysics, and heliophysics should work closely with EPRV survey teams to model absorption line profiles (Wright and Sigurdsson, 2018). Observing the Sun simultaneously with spatially resolved spectroscopy and in hemisphere-integrated light (e.g., Haywood et al., 2016; Dumusque et al., 2015) offers the opportunity to uniquely identify the surface features that result in apparent RV variations, and thus a path for comparing theoretical calculations to disk-integrated measurements.* (p.93)

and

> *in order to achieve the RV precision to detect an Earth-mass planet orbiting a Sun-like star, increased observational cadence, spectral resolution, SNRs, and stellar activity monitoring are required to disentangle stellar from planetary signals. This approach should include close collaboration with the solar and stellar astrophysics communities, including theorists, modelers, and observers (Wright and Sigurdsson, 2018, white paper).* (p.97)